\documentclass[square,aip,reprint,showkeys,groupedaddress]{revtex4}
\usepackage{graphicx}
\begin{document}

\title{Graphical Analysis of Current-Voltage Characteristics in Typical Memristive Interfaces}

\author{C. Acha}
\thanks{corresponding author (acha@df.uba.ar)}
\address{Laboratorio de Bajas
Temperaturas - Departamento de F\'{\i}sica - FCEyN - Universidad de
Buenos Aires and IFIBA - CONICET,  Pabell\'on I, Ciudad
Universitaria, C1428EHA Buenos Aires, Argentina}

\date{\today}


\begin{abstract}

A graphical representation based on the isothermal current-voltage
(IV) measurements of typical memristive interfaces is presented.
This is the starting point to extract relevant microscopic
information of the parameters that control the electrical properties
of a device based on a particular metal-oxide interface. The
convenience of the method is illustrated presenting some examples
were the IV characteristics were simulated in order to gain insight
on the influence of the fitting parameters.

\end{abstract}



\maketitle

\section{Introduction}

The study of the current-voltage (IV) characteristics can be a very
useful tool to understand which are the microscopic factors that
determine the main conduction mechanism through a metal-oxide
interfaces of a device.~\cite{Sze06,Chiu14} This microscopic
knowledge may be a route to improve their capacities and a clever
way to modify some specific properties. In the particular case of
memristors or devices based on the resistive switching (RS)
properties~\cite{Waser07,Strukov08,Sawa08}, which give rise to
non-volatile memories called Resistive Random Access Memories
(RRAM), to understand if the resistance state (high or low) depends
on electrode's or bulk's microscopic properties is an essential task
in order to design their functionalities. In the case of the
electrode-limited devices, the work function of the metal, the
carrier affinity and the thickness of the oxide determine the
barrier height and the probability to produce an
electric-field-induced-current through the junction. The conduction
mechanism can then be described as Schottky (Sch), Fowler-Nordheim
(FN) or as direct tunneling emission. While in the case of
bulk-limited interfaces the conduction mechanism is determined by
the electrical properties of the oxide near the interface as, for
example, those imposed by the existence of traps and their energy
levels. Two examples of transport mechanisms influenced by the
energy distribution and density of traps are the Poole-Frenkel
emission (P-F) and the space-charge-limited conduction (SCLC). As an
example of a device based on an electric-field-trap-controlled SCLC
mechanism we can mention Ag/La$_{0.7}$Ca$_{0.3}$MnO$_{3-\delta}$
interfaces~\cite{Shang06}, while Au/YBa$_2$Cu$_3$O$_{7-\delta}$
interfaces show a PF conduction in a variable-range hopping
scenario, with a pulse-controlled-trap energy level that determines
their resistance switching properties.~\cite{Schulman15}

\begin{figure} [!t]
\centering \vspace{0mm}
\includegraphics[width=9cm]{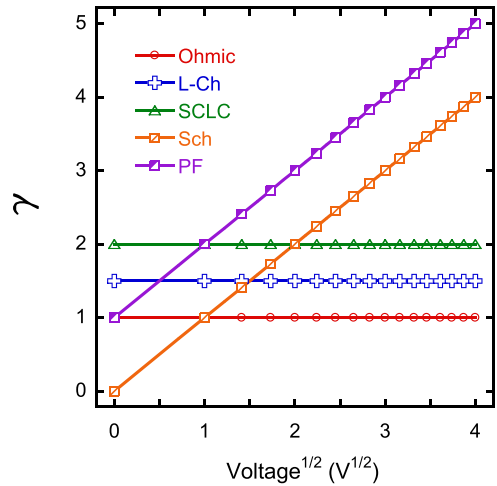}
\vspace{-0mm} \caption{(Color online) The power exponent  $\gamma=
dLn(I)/dLn(V)$ vs $V^{1/2}$ representation of some typical
conduction processes through metal-oxide interfaces: ohmic,
Langmuir-Child (L-Ch), space charge limited currents (SCLC),
Schottky (Sch) and Poole-Frenkel (PF).} \vspace{-0mm}
\label{fig:puros}
\end{figure}

In this way, different scenarios can be considered to explain the
microscopic origin of the resistance change of the device, related
to the particular choice of materials for the metal/oxide interface.
As shown previously~\cite{Sze06,Chiu14}, by using their isothermal
IV characteristics that it is possible to distinguish if the
conduction of the device is related to an ohmic behavior ($I \sim
V$), or a space charge limited conduction (SCLC, $I \sim V^2$), or a
Poole-Frenkel (PF), Fowler-Nordheim (FN) or Schottky (Sch) emissions
[$I \sim \exp(V^n)$]. A simple way to enlighten the origin of the
main conduction mechanism of a device is to plot as a function of
$V^{1/2}$ the power exponent parameter $\gamma$, defined as $\gamma
= dLn(I) / dLn(V)$, instead of trying to fit by trail and error
their IV characteristics.~\cite{Bozhko02} This is because the
typical conduction mechanisms through metal-oxide interfaces have a
simple $\gamma$ vs $V^{1/2}$ curve (see Fig.~\ref{fig:puros}): a
pure ohmic, a Langmuir-Child (L-Ch) or a SCLC conduction will show a
constant $\gamma$ ($=$ 1, 1.5 or 2, respectively), and a Schottky
(Sch) or a Poole-Frenkel (PF) behavior will be represented by a
straight line, only differing in the intercept (0 for Sch, 1 for
PF).~\cite{Sze06}

The use of this power exponent representation has proved to be
convenient for extracting physical parameters in the case of
non-ideal diodes.~\cite{Mikhe99} Here, we show that if a combination
of conducting mechanism are present in a particular metal-oxide
interface, this representation would be very useful to graphically
determine what they are. Otherwise, the standard way to do it will
require to analyze the qualities of the fits to the IV
characteristics of multiple expressions to test. It should be noted
that a good fit of the IV characteristics, which includes all the
participating mechanisms, is essential in order to extract a
flawless microscopic description of the physics behind these
devices.

By considering previous works~\cite{Acha11,Marlasca13,Acha16}, the
combination of conduction mechanism seems to be a common feature of
bipolar memristive interfaces, as a non-linear element in parallel
and/or in series with an ohmic resistance gives a convenient
representation. The physical explanation of the microscopic origin
of the parallel and the series structure is not clear yet. We may
speculate that it can be a consequence of their capacitor-like
structure where both a phase-separated interfacial zone and the
proper oxide have a relevant participation in the conducting
process. The interfacial zone would be composed by a mixture of
conducting and insulating
regions~\cite{Lunkenheimer02,Schulman15,Acha16}, probably associated
with a disordered distribution of oxygen vacancies (phase
separation), leading to the existence of a non-linear element in
parallel with an ohmic one. The series ohmic element would then
represent the bulk contribution of the oxide.

\section{The power exponent $\gamma$ representation}

\begin{figure} [!t]
\centering \vspace{0mm}
\includegraphics[width=10cm]{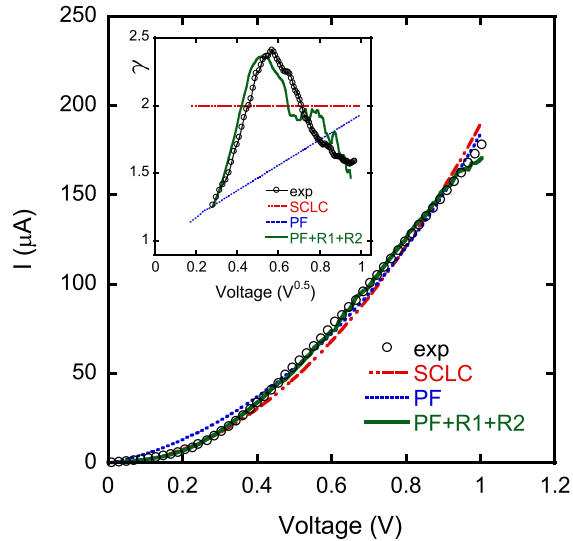}
\vspace{-0mm} \caption{(Color online) IV characteristics of a
Ag-cobaltite interface. Lines are fits considering a PF or a SCLC
non-linear conduction and a combination of a PF conduction in
parallel with an ohmic element, both in series with a second ohmic
element. This latter combination of conduction processes was
determined by the non-trivial $\gamma$ representation of the
experimental data (shown in the inset). For comparison, the $\gamma$
values of the fitting models are also represented in the inset. }
\vspace{-0mm} \label{fig:ejemplo}
\end{figure}

The convenience of the $\gamma$ representation can be illustrated by
the example shown in Fig.~\ref{fig:ejemplo}, corresponding to a
typical IV characteristic of a Ag-La$_{0.7}$Sr$_{0.3}$CoO$_3$
interface. Experimental details of the samples, the experimental
setup and the way the measurements were performed can be found
elsewhere.~\cite{Acha16} A Sch mechanism can be ruled out as non
rectifying behavior was observed: although not shown here, there was
no appreciable difference in the behavior of the data when comparing
between the negative and the positive quadrant.

The IV characteristics were fitted using the expressions of pure PF
and SCLC conduction mechanisms~\cite{Sze06,Chiu14} (see the solid
lines in Fig.~\ref{fig:ejemplo}). Fits are not good and a
combination of conduction mechanism should be considered to give a
right representation of the data. At this point, the way to
determine which are the relevant conducting processes is somehow
tedious as it requires to fit the IV data considering, by trial and
error, each possible combination of the electrical conduction
processes (PF or SCLC and an ohmic resistance in parallel, or with
an ohmic resistance in series, or with both of them). Hopefully, the
$\gamma$ representation of the experimental data gives a quick
graphical solution to determine the specific mechanism involved, as
can be observed in the inset of Fig.~\ref{fig:ejemplo}. The $\gamma$
representation of the pure PF and SCLC conduction mechanism are
clearly far from the experimental data. As we will show later, the
existence of a cusp with values of $\gamma
>2$ with the tendency to reach an asymptotic value of 1 both in the
low and high voltage regions are clear evidences of a PF conduction
in parallel with an ohmic element, both in series with a second
ohmic element. Then, as also shown in Fig.~\ref{fig:ejemplo}, the IV
experimental data was nicely fitted with the corresponding
mathematical expression to those processes. It can be seen that the
$\gamma$ representation of the fitting data also follows the
experimental $\gamma$ curve. In the next section we will discuss in
detail how the $\gamma$ representation differs in some selected
examples.

\subsection{Circuit representation for bipolar memristors}

\begin{figure} [!t]
\centering \vspace{0mm}
\includegraphics[width=4cm]{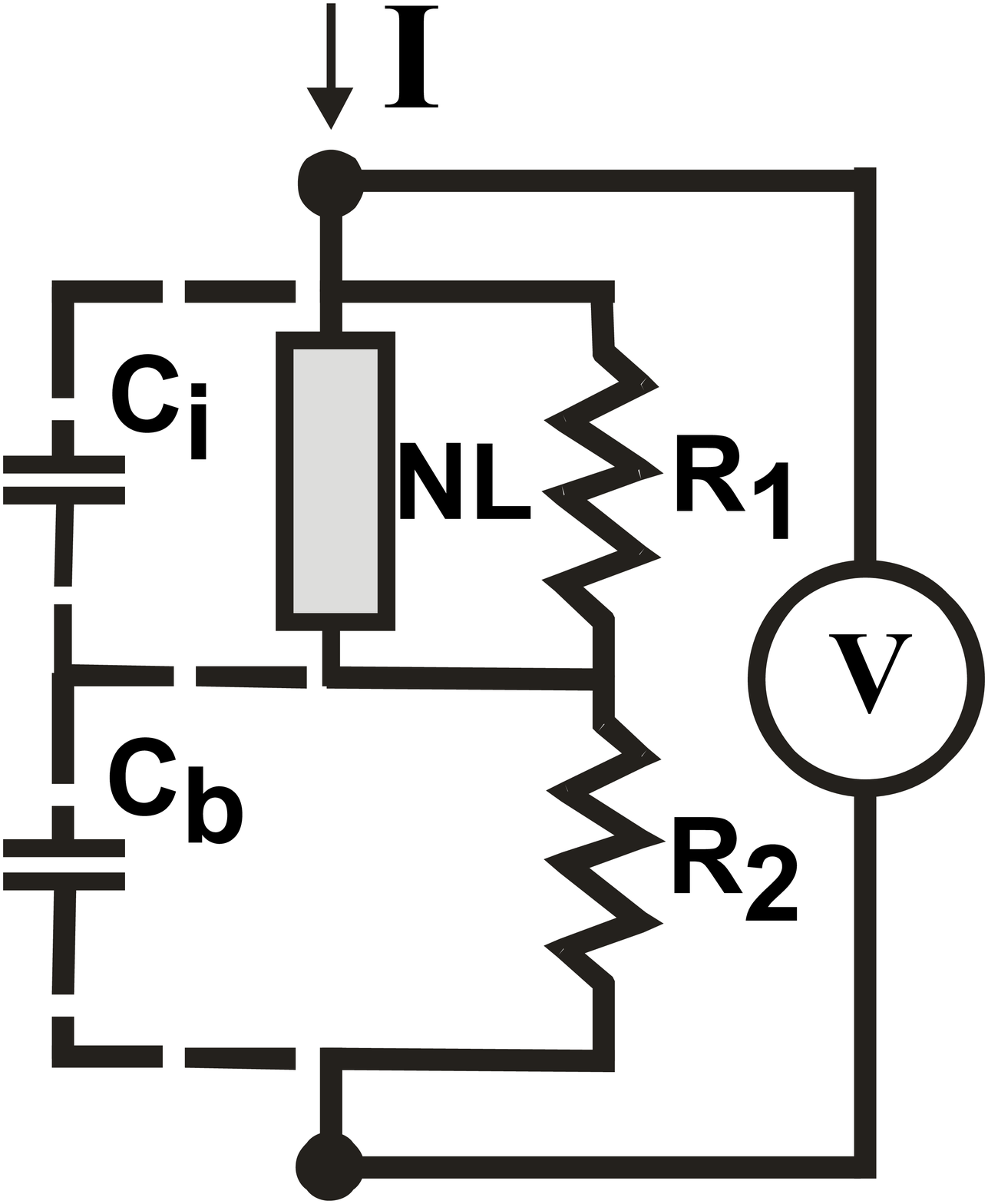}
\vspace{-0mm} \caption{(Color online) Circuit elements proposed to
represent the IV characteristics of bipolar memristors. A non-linear
element (NL) and an ohmic resistance $R_1$ represents the
interfacial elements while a second ohmic resistance $R_2$ in series
represents the bulk. The interfacial ($C_i$) and bulk ($C_b$)
capacitances have been included for completitude with dashed lines
as their contribution is not relevant here because we are only
considering the static response to electric fields.} \vspace{-0mm}
\label{fig:circuito}
\end{figure}

In order to gain insight into the $\gamma$ representation, we
propose to analyze the IV characteristics of the circuit elements
shown in Fig.~\ref{fig:circuito}. The interfacial resistance is
simulated by a non-linear element (NL) in parallel with an ohmic
resistance $R_1$. The NL element can be based on the large variety
of conduction mechanism through a metal-oxide interface described in
the previous section. Here, we will analyze two particular cases of
NL element (SCLC and PF) which are common for bipolar memristors
although the same analysis can be extended for devices dominated by
a Sch conduction. The capacitance of this zone, related to the
dielectric constant of the interface~\cite{Lunkenheimer02}, was not
considered here, as we are focusing our description on the low
frequency regime. Other circuit representations of memristors can be
found elsewhere.~\cite{Blasco14,Blasco15,Radwan15,Vourkas16} The
convenience of this circuit to describe the behavior of memristive
interfaces was tested by reproducing the dynamical behavior of
metal-YBCO interfaces~\cite{Acha11} as well as by capturing the
non-trivial IV characteristics of metal-manganite
junctions~\cite{Marlasca13}. In the case of devices with
non-negligible bulk resistance, a second ohmic element in series
($R_2$) should be considered.

\subsection{The SCLC mechanism as NL element}

We first analyze the simple case of an SCLC element without traps
(Child's law)~\cite{Mark62} in parallel with an ohmic resistance
$R_1$. In this case the current ($I$) dependence on applied voltage
($V$) corresponds to the following equation:

\begin{equation}
\label{eq:SCLCconR1} I = A ~V^2 + \frac{V}{R_1},
\end{equation}

\noindent with $A = \frac{9}{8} \frac{\mu \epsilon}{d^3}$, where
$\mu$ is the carrier mobility, $\epsilon$ the static dielectric
constant and $d$ the distance between contacts.

\begin{figure} [!h]
\centering \vspace{0mm}
\includegraphics[width=9cm]{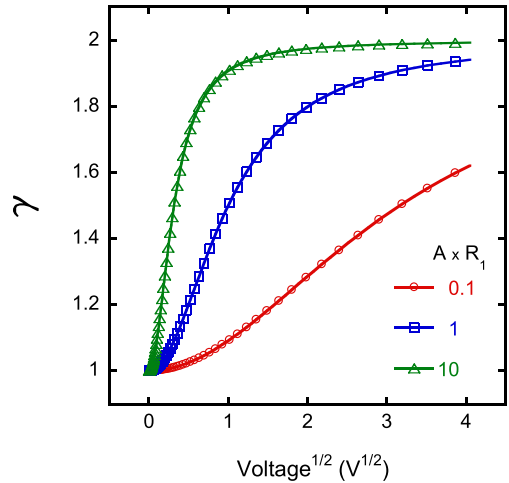}
\vspace{-0mm} \caption{(Color online) $\gamma$ representation for a
SCLC conduction with an ohmic element $R_1$ in parallel. The $A$ x
$R_1$ factor controls the voltage dependence of $\gamma$ from an
ohmic ($\gamma = 1$) to a SCLC ($\gamma = 2$) behavior.}
\vspace{-0mm} \label{fig:SCLC paral}
\end{figure}

As shown in Fig~\ref{fig:SCLC paral}, $\gamma$ increases
asymptotically from 1 to 2 depending on the $A$ x $R_1$ factor. The
saturation to 2 is obtained at lower voltages when increasing this
factor.

Secondly, we consider the SCLC element only in series with the ohmic
element $R_2$. In this case the current $I$ is related to the total
voltage drop $V$ by

\begin{equation}
\label{eq:SCLCconR2} I = A ~(V-IR_2)^2,
\end{equation}

This second degree equation can be solved to extract the relation
between I and V in order to calculate the $\gamma$ factor. As shown
in Fig.~\ref{fig:SCLC serie}, $\gamma$ evolves from the pure SCLC
value 2 at low voltages to the ohmic dependence 1 at higher
voltages. The velocity of this evolution increases when increasing
the $A$ x $R_2$ factor.

\begin{figure}
\vspace{0mm}
\includegraphics[width=9cm]{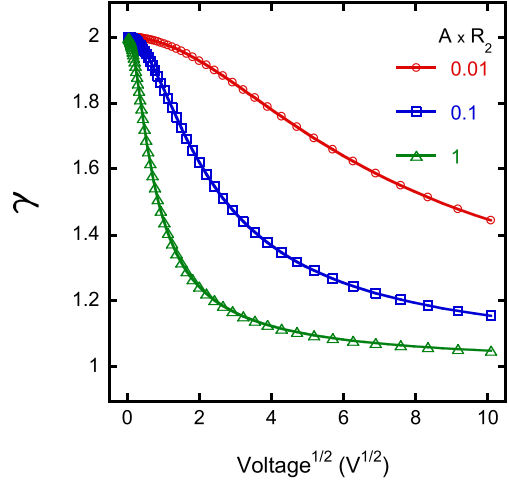}
\vspace{-0mm}\caption{(Color online) $\gamma$ representation for a
SCLC conduction with an ohmic element $R_2$ in series. Here, the $A$
x $R_2$ factor regulates the voltage sensitivity of $\gamma$ that
evolves, with increasing the voltage, from a SCLC ($\gamma = 2$) to
an ohmic ($\gamma = 1$) behavior.} \vspace{-0mm} \label{fig:SCLC
serie}
\end{figure}

\begin{figure} [!h]
\vspace{0mm}
\includegraphics[width=9cm]{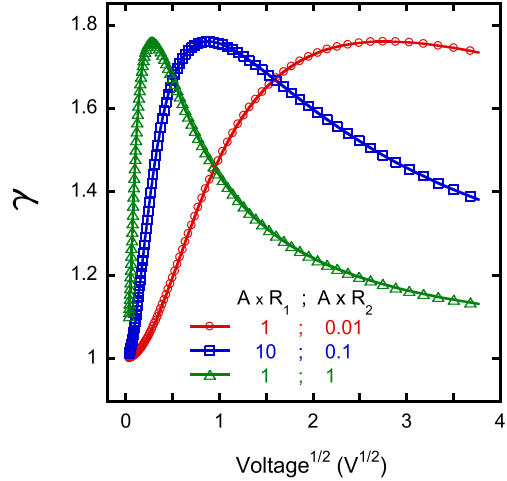}
\vspace{-0mm}\caption{(Color online) $\gamma$ representation for a
SCLC conduction with an ohmic element $R_1$ in parallel and a second
ohmic element $R_2$ in series. A peak can be observed, where the
factors $A$x$R_1$ determines its position and $A$x$R_2$ its width.}
\vspace{-0mm} \label{fig:SCLC paral_serie}
\end{figure}

Finally, when both $R_1$ and $R_2$ are present, the relation between
I and V is also obtained by solving the Eq.~\ref{eq:SCLCconR1yR2}. A
superposition of both previous behaviors can be observed in
Fig.~\ref{fig:SCLC paral_serie}. Here, the $A$ x $R_1$ controls the
voltage where the $\gamma$ peak is obtained (always with $\gamma
\leq 2$), while $A$ x $R_1$ determines the rapidity to tend to an
ohmic behavior when increasing the voltage.

\begin{equation}
\label{eq:SCLCconR1yR2} I = A ~(V-IR_2)^2 + \frac{(V-IR_2)}{R_1},
\end{equation}

\subsection{The PF mechanism as NL element}

When dealing with a NL element related to a PF mechanism in a
circuit that includes both $R_1$ and $R_2$ (see
Fig.~\ref{fig:circuito}), the relation between the current $I$ and
the voltage $V$ corresponds to the following implicit equation:

\begin{equation}
\label{eq:PFconR1R2a} I = (\frac{V-IR_2}{R_1})\{\frac{R_1}{R_{PF}}
\exp[\frac{C(V-IR_2)^{1/2}}{k_B T}] + 1\},
\end{equation}

\noindent with
\begin{equation}
\label{eq:PFconR1R2b} R_{PF} = \frac{R_{ox}}{\exp(-\frac{\phi_B}{k_B
T})}, C = \frac{q^{3/2}}{(\pi \epsilon^{'} d)^{1/2}}, R_{ox} =
\frac{d}{Sqn_0\mu},
\end{equation}

\noindent where $T$ is the temperature, $k_B$ the Boltzmann
constant, $\phi_B$ the trap energy level, $q$ the electron's charge,
$S$ the conducting area, $\epsilon^{'}$, $n_0$ and $ \mu$ the real
part of the dielectric constant, the density of carriers and their
mobility in the oxide, respectively, $d$ the distance where the
voltage drop is produced and $R_{PF}$ the resistance of the PF
element when $V\mapsto0$.

\begin{figure}
\vspace{0mm}
\includegraphics[width=9cm]{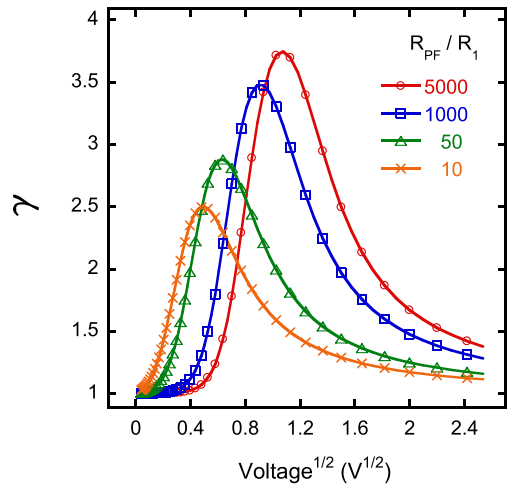}
\vspace{-0mm}\caption{(Color online) $\gamma$ representation for a
PF element in parallel with an ohmic resistance ($R_1$) and in
series with a second resistance ($R_2$). The $\gamma$ curves are
plotted for different $R_{PF}/R_1$ ratios. } \vspace{-0mm}
\label{fig:PFR1R2}
\end{figure}

\begin{figure}
\vspace{0mm}
\includegraphics[width=9cm]{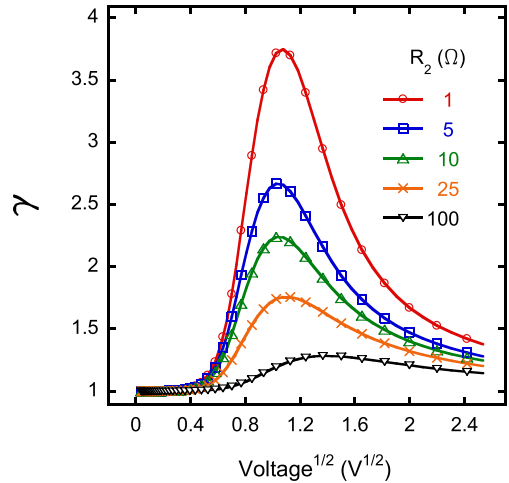}
\vspace{-0mm}\caption{(Color online) $\gamma$ representation for a
PF element in parallel with an ohmic resistance ($R_1$) and in
series with a second resistance ($R_2$). The $\gamma$ curves are
plotted for different $R_2$ values. } \vspace{-0mm}
\label{fig:PFvarR2}
\end{figure}

\begin{figure}
\vspace{0mm}
\includegraphics[width=9cm]{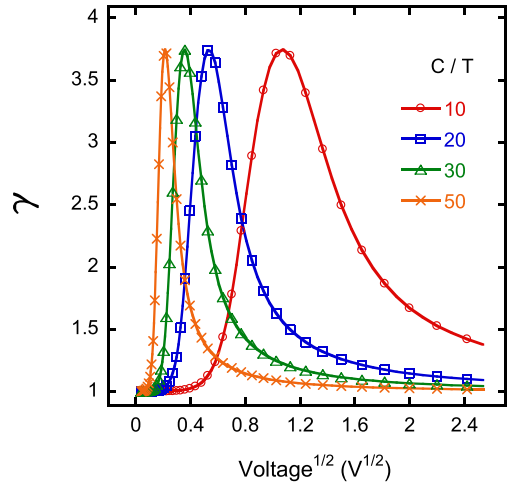}
\vspace{-0mm}\caption{(Color online) $\gamma$ representation for a
PF element in parallel with an ohmic resistance ($R_1$) and in
series with a second resistance ($R_2$). The $\gamma$ curves are
plotted for different $C/T$ ratios.} \vspace{-0mm}
\label{fig:PFvarCT}
\end{figure}

This equation should be solved numerically in order to determine the
IV curves and to calculate the $\gamma$ factor. As can be derived
from Eq.~\ref{eq:PFconR1R2a}, three different factors control the IV
dependencies: $\frac{R_{PF}}{R_1}$, $R_2$ and $\frac{C}{T}$. The
influence of each factor was analyzed (while maintaining constant
the others) as shown in Fig.~\ref{fig:PFR1R2}, \ref{fig:PFvarR2} and
\ref{fig:PFvarCT}.

As can be observed in Fig.~\ref{fig:PFR1R2}, the ohmic region at low
voltages increases when increasing the ratio $\frac{R_{PF}}{R_1}$.
This fact is associated with the weight in the conduction process of
each element, so that the ohmic region delays the PF contribution
when $R_1 \ll R_{PF}$. With regard to the peak of $\gamma$, it is
developed in the region of influence of the PF element and may reach
values $>$ 2. In this way, if an increasing $\gamma > 2$ is observed
when analyzing the $\gamma$ plots of experimental data, the
existence of an SCLC mechanism (Child's law) as the main NL element
can be ruled out.

When $R_1$ $\leq$ $R_{PF}$, $R_1$ becomes short circuited by the PF
element when increasing the voltage and the overall behavior becomes
asymptotically ohmic by the main contributions to the IV
characteristics imposed by $R_2$.

When increasing $R_2$, the voltage region of the PF element
influence is decreased in a way that the $\gamma$ peak is reduced
(eventually with values $<$ 2) and becomes wider, making it barely
visible, as depicted in Fig.~\ref{fig:PFvarR2}

Finally, as shown in Fig.~\ref{fig:PFvarCT}, when the ratio
$\frac{C}{T}$ increases, the voltage width of the $\gamma$ peak,
that indicates the extension of the PF influence in the circuit
behavior, is reduced. This figure points out the importance of
performing IV measurements at different temperatures (see
ref.\cite{Acha16}), as this can be a way to determine (or to check)
the temperature dependence of microscopic parameters, like
$\epsilon^{'}$(T) (from $C$) or $\phi_B$ (from the temperature
dependence of $R_{PF}$), as well as of other parameters, such as
$R_1$ or $R_2$, which can also be temperature dependent.

It is interesting to note that when performing the fits to the
experimental data, the number of fitting parameters can be reduced
if the low voltage resistance of the device is measured (usually
called $R_{rem}$). For example, in the case of the PF element with
the ohmic resistances $R_1$ and $R_2$, by analyzing
Eq.~\ref{eq:PFconR1R2a} in the low voltage limit, it can be shown
that $R_2 \simeq R_{rem} - (1/R_{PF} + 1/R_1)$.

These examples show the utility of the $\gamma$ representation when
dealing with the experimental curves measured at different
temperatures: just by graphically analyzing the shape of this curve,
the existence of a particular NL element can be determined as well
as the necessity to include additional ohmic elements as $R_1$ and
$R_2$. Then, the proper IV relation can be used in order to fit the
data and to extract the microscopic parameters that govern the
conduction mechanism of a device.

\vspace{-0mm}

\section{Conclusion}
We have shown that the power exponent parameter $\gamma$ plotted as
a function of $V^{1/2}$ can be a useful tool to graphically
determine the conduction mechanisms through an interface. This
method becomes particularly interesting when the contribution to the
conduction process comes from a combination of different elements,
including a NL element in series and/or in parallel with ohmic ones.
As this is the typical scenario found for some memristive
interfaces, the idea is to ease the determination of these elements
in order to obtain relevant microscopic information by fitting their
IV characteristics with the corresponding expression.

\section{Acknowledgment}

We would like to thank the financial support by CONICET Grant PIP
112-200801-00930, PICT 2013-0788 and UBACyT 20020130100036BA
(2014-2017). We are grateful to A. Schulman, M. Boudard, K. Daoudi
and T. Tsuchiya for providing the experimental data shown in Fig.2.
We also acknowledge Dr. V. Bekeris for a critical reading of the
manuscript, and D. Gim\'enez, E. P\'erez Wodtke and D.
Rodr\'{\i}guez Melgarejo for their technical assistance.


\end{document}